\documentclass[journal,draftcls,onecolumn,12pt,twoside]{IEEEtran}
\usepackage[latin1]{inputenc}
\usepackage[english]{babel}
\usepackage{amsmath,amsfonts,amssymb}
\usepackage{graphicx, epsfig}
\usepackage{subfigure}
\usepackage{array}
\usepackage{multirow}
\usepackage{mdwmath}
\usepackage{mdwtab}
\usepackage{algorithmic}
\usepackage{algorithm}
\usepackage{color}

\usepackage{color}

\ifCLASSINFOpdf

\else

\fi

\begin{document}
%
\title{\huge Coding Techniques for Future Wi-Fi}

\author{\IEEEauthorblockN{Andre G. D. Uchoa and Rodrigo C. de Lamare$^{*}$}
\IEEEauthorblockA{\\CETUC - Pontifical Catholic University of  Rio de Janeiro - PUC-RJ,
R. Marques de Sao Vicente, 225 - Ala Kennedy, 22451-900, Gavea, RJ, Brazil\\
Department of Electronics, University of York, UK$^{*}$\\
$\lbrace$agdu, delamare$\rbrace$@cetuc.puc-rio.br}}

\maketitle

\begin{abstract}
Future wireless applications such as high definition video
streaming, wireless cloud radio access networks, and cellular data
offload are bandwidth-hungry, which highlights the need for Wi-Fi
links exceeding 1 Gb/s. This article discusses coding schemes and,
in particular, the use of Low-Density Parity-Check (LDPC) codes as
Forward Error Correction (FEC) in future Wi-Fi standards. Moreover,
we consider advanced design strategies such as root-check LDPC
structures, computer-aided design for short blocks and
high-performance decoding algorithms with low latency. We then
conclude the article with a discussion of FEC challenges for future
Wi-FI applications.
\end{abstract}


%
\IEEEpeerreviewmaketitle

\section{Introduction}

Driven by the ever-increasing user connectivity demands, emerging
and future generations of Wi-Fi, such as IEEE 802.11ac and WiGig
IEEE 802.11ad, will be capable of achieving multiple gigabits per
second speeds. Furthermore, the future Wi-Fi will be used to do
everything from simple web browsing and peer-to-peer sharing, to
multimedia streaming, real-time teleconferencing, cable replacement,
and wireless docking, to name a few. Therefore, one key element will
be the reliable delivery of information at the final destination.
One strategy to improve the reliability of the received information
is the use of Forward Error Correction (FEC). FEC relies on a
mathematical mapping of messages to include a special kind of
redundancy that enables the receiver to correct the errors caused by
the channel \cite{ryanbook}.

There are two main types of FEC used by Wi-Fi: LDPC codes and
convolutional codes. LDPC codes belong to the class of block codes,
can perform close to channel capacity and achieve excellent
performance. Block codes work on fixed-size blocks (packets) of bits
or symbols of predetermined size. A practical block code can be
decoded in polynomial time to its block length. In contrast,
convolutional codes work on bit or symbol streams of arbitrary
length. Convolutional codes are most often decoded with the Viterbi
algorithm \cite{ryanbook}, though other algorithms are sometimes
used. Nevertheless, the Viterbi algorithm can lead to high
complexity in terms of decoding for codes with large constraint
length.  Recent studies comparing convolutional and LDPC codes have
reported that convolutional codes can be advantageous for low
latencies, whereas LDPC codes are preferred for higher latencies or
medium to large blocks \cite{maiya.12}.

\begin{figure*}[htb]
 \centering
\includegraphics[scale=0.55]{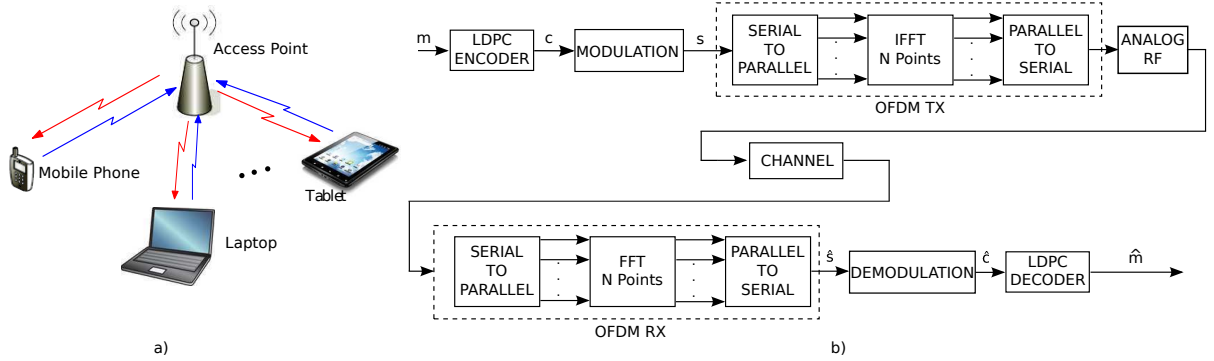}
\caption{Simplified system model for a WLAN with Wi-Fi. a) An example of scenario with a tablet, a laptop, a mobile phone and an access point. b) The transmitter and receiver structure of an LDPC-coded Wi-Fi system.} \label{fig:fig1}
\end{figure*}

In Fig. \ref{fig:fig1} an example of a system model for a Wireless
Local Area Network (WLAN) is presented. In Fig \ref{fig:fig1} a), we
have a Wi-Fi communications system where a tablet, a laptop and a
mobile phone exchange information with an Access Point (AP). In Fig.
\ref{fig:fig1} b) a simplified structure of a Wi-Fi communication
system between a transmitter (TX) and a receiver (RX), for example,
a laptop and an AP is shown. In Fig. \ref{fig:fig1} b) we describe
the encoding, transmit and receive processing, and the decoding of a
message ${\bf m} = [m_o \ldots m_k]$. Specifically, the message
${\bf m}$ is encoded by an LDPC encoder to produce the codeword
${\bf c} = [c_o \ldots c_n]$, where the rate is $R = k/n$. Then, the
codeword ${\bf c}$ is modulated to one of the available schemes,
namely, BPSK, QPSK, 16-QAM, 64-QAM and 256-QAM generating the symbol
stream $\mathbf{s}$. In the case of IEEE 802.11ac the symbol stream
$\mathbf{s}$ is transmitted by an OFDM technique which involves
additional steps: a serial to parallel ($1$ up to $8$ streams)
conversion to calculate the Inverse Fast Fourier Transform (IFFT)
with a maximum of 114 sub-carries/pilots, a parallel to serial
conversion is performed, and the resulting stream is sent to the
channel. At the receiver side the inverse process is carried out.
For the case of IEEE 802.11ad the transmission is omitted (dashed
blocks), so the resulting stream can be transmitted through either
an OFDM scheme or a single-carrier one. Extensions to OFDM systems
\cite{gfdm,uwgfdm} and multiple-antenna systems \cite{wence} can
also be considered. The focus of this article is on LDPC design and
decoding strategies for future Wi-Fi.

Following the system model of Fig. \ref{fig:fig1} b) a ${\bf G}$
generator matrix to encode the message ${\bf m}$ is used and a
Parity Check Matrix (PCM) ${\bf H}$ for the decoding process. If the
estimated codeword $\hat{\bf c}$ is correct then $\hat{\bf c} \cdot
{\bf H}^{T} = 0$ must be satisfied, where $(.)^{T}$ stands for the
transpose operation. An LDPC code is characterized by its sparse PCM
${\bf H}$. For ${\bf H}$ to be sparse the number of entries equal to
one must be much less than the number of zeros.

%


In this paper we discuss the perspectives of LDPC coding for the
future of Wi-Fi. Furthermore, the LDPC codes used in the Wi-Fi
standard are reviewed. Moreover, we draw attention to advanced
design strategies such as Root-Check LDPC structures, computer-aided
methods for short blocks and how they can be used in future
standards. In addition, we also consider advanced LDPC decoding
techniques such as reweighting and scheduling methods, which have
the potential to reduce the decoding latency in several important
applications. An analysis of fast LDPC decoding methods to decrease
the computational complexity in hardware is also included. Finally,
some research challenges are identified.

The rest of this paper is organized as follows. In Section II a
general discussion about LDPC codes for wireless communication
standards is developed. In Section III we describe the LDPC codes
used in the Wi-Fi standard. In Section IV we review the use of
Root-Check LDPC codes and other techniques for the future Wi-Fi. In
Section V we discuss advanced decoding techniques such as
reweighting, scheduling and reduced complexity decoding methods.
Section VI presents the challenges, while Section VII concludes this
paper.

\section{LDPC Codes in Wireless Communication Standards}

LDPC codes with randomly generated parity check matrices and large
blocks generally have good performance, but they lack enough
structure to facilitate efficient encoding methods. In practical
applications, structured and short length LDPC codes are highly recommended as
modern wireless standards adopt Quasi-Cyclic LDPC
(QC-LDPC) codes and Irregular Repeat and Accumulate LDPC (IRA-LDPC)
codes.

QC-LDPC codes are designed by tiling circulant matrices. A circulant
matrix is a square matrix in which each row is a right cyclic shift
of the previous row and the first row is a cyclic shift of the last
row. A circulant matrix can be totally characterized by its first
row. If a circulant matrix has weight $w = 1$ per row, it is called
a circulant permutation matrix. The all-zero matrix, also called
null matrix, is also circulant matrix with all its elements being
zero. If a PCM ${\bf H}$ consists of $m \times n$ circulant
sub-matrices of dimensions $s \times s$ with $M = ms$ and $N = ns$,
the resulting linear block code will be a QC code with a period of
$n$. In such code, the $n$-bit shift of any codeword is another
codeword. The generator matrices ${\bf G}$ of QC-LDPC codes are in
systematic-circulant form with the requirement that the PCM are full
rank. The memory cost for encoding QC-LDPC codes is greatly reduced
and the encoder can be implemented by using simple shift registers
\cite{ryanbook}.

\begin{figure}[htb]
 \centering
\resizebox{128mm}{!}{
\includegraphics{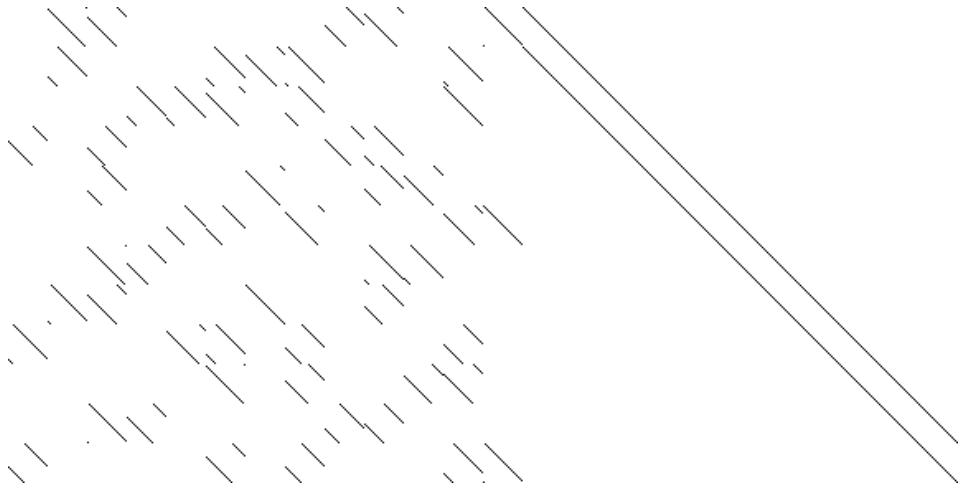}
}\caption{An example of a parity check matrix for a QC-IRA-LDPC code.
The black dots represent the ones and the white space
represent the zeros.} \label{fig:pcm_qcira_r12}
\end{figure}

The accumulator-based codes that were invented first are the
so-called repeat-accumulate (RA) codes \cite{ryanbook}. Despite
their simple structure, they were shown to provide good performance
and, more importantly, they pioneered the design of efficiently
encodable LDPC codes. The key points in using IRA-LDPC codes are the
simplicity in designing such codes and faster encoding than
conventional LDPC methods. The Wi-Fi standard uses a type of QC-LDPC
codes, whereas other standards work with a combination of QC and IRA
LDPC codes known as QC-IRA LDPC codes. There are two main advantages
in QC-IRA-LDPC codes: the memory requirements to store the matrices
by QC codes and the simple encoding process provided by IRA codes.
For instance, Fig. \ref{fig:pcm_qcira_r12} presents an example of a
PCM for a QC-IRA-LDPC code with dimension $288 \times 576$. The right
hand side of Fig. \ref{fig:pcm_qcira_r12} is a dual diagonal associated with an RA based code. On the left hand side of
Fig. \ref{fig:pcm_qcira_r12} are the circulant matrices.

\section{LDPC Codes Used In The Wi-Fi Standard}

In this section, the LDPC codes used in the Wi-Fi standard are
detailed. The Wi-Fi standard adopts Quasi-Cyclic LDPC (QC-LDPC)
codes. As discussed previously, FEC introduces some kind of
redundancy in the form of $n-k$ extra bits in a codeword of length $n$
to improve the reliability of a delivered message which results in
the code rate $R=k/n$. For instance, the PCM presented on Fig.
\ref{fig:pcm_qcira_r12} is an LDPC code with code rate $R =
\frac{1}{2}$. The rate $\frac{1}{2}$ means that half of the codeword
is the message (information) and the other half is the redundancy.
The Wi-Fi standard operates with the following code rates, $R =
\frac{1}{2}$, $\frac{3}{4}$, $\frac{5}{8}$ and $\frac{13}{16}$ with
a fixed block length of $L = 672$ bits. For example, with a code
rate $\frac{1}{2}$ we have $336$ information bits plus $336$ bits of
redundancy, for code rate $\frac{13}{16}$ we have $546$ information
bits plus $126$ bits of redundancy.

\begin{figure}[htb]
 \centering
\resizebox{128mm}{!}{
\includegraphics{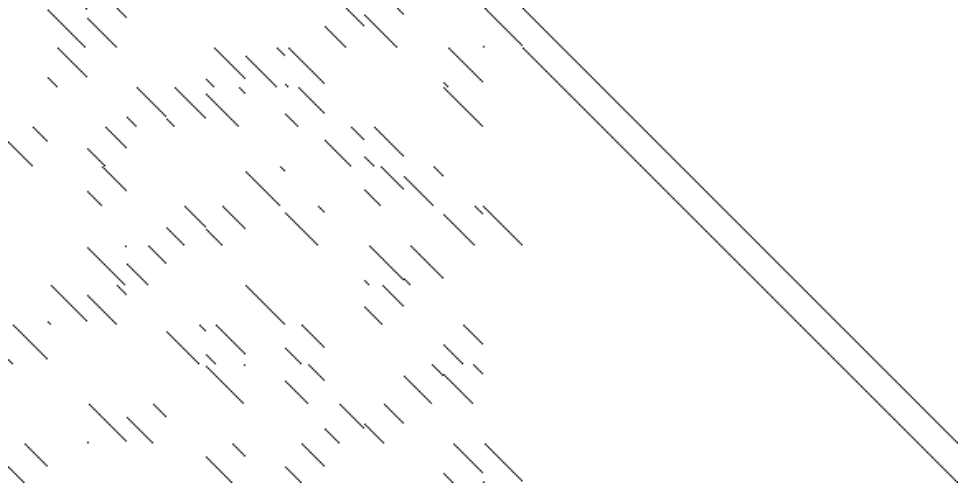}
}\caption{The PCM $\mathbf{H}$ for code rate $R = \frac{1}{2}$ from Wi-Fi standard. The black dots are the ones and the white space represent the zeros.} \label{fig:pcm_wifi_r12}
\end{figure}

Fig. \ref{fig:pcm_wifi_r12} presents the PCM ${\bf H}$ of the Wi-Fi
standard with code rate $R = \frac{1}{2}$ and dimensions $336 \times
672$. The other PCMs for the rest of code rates follow a similar
structure as for code rate $R = \frac{1}{2}$. Once a standard is
defined, a design is adopted and the PCM ${\bf H}$ must be stored in
all devices for compatibility issues. Whenever the standard is
updated there is an opportunity to incorporate more sophisticated
PCM designs which could lead to improved performance.

\section{Designing High-Performance LDPC codes}

In this section we review advanced LDPC code design strategies such
as Progressive Edge Growth (PEG) algorithms, Root-Check LDPC
structures, QC and IRA structures.

\subsection{PEG-Based Algorithms}

Among the algorithms capable of producing LDPC codes of highest
performance and efficiency for short to moderate lengths are the
PEG-based algorithms \cite{hu.05} and QC and IRA structures. The
codes produced by the PEG algorithm exhibit improved performance
compared to random construction methods \cite{hu.05} as PEG
optimizes the underlying graph structure of the PCM. In particular,
PEG strategies attempt to increase the girth of the graph and
improve its general connectivity. Improvements and modifications in
the original PEG algorithm have provided a better performance than
the original PEG algorithm. For instance, Healy and de Lamare in
\cite{dopeg.comms.12} have proposed an algorithm to design LDPC
codes which involves the use of decoder-based optimisation with the
sum-product algorithm (SPA) and also considered multiple candidates
to optimize the graph connections \cite{memd}. Therefore, LDPC codes
for future Wi-Fi standards can take advantage if they are designed
by PEG-based algorithms and several detection and decoding
strategies \cite{spa,tds,mbdf,did,bfidd,1bitidd}.

\subsection{Structured Code Designs}

A drawback of standard unstructured LDPC codes designed with
conventional and PEG-based algorithms is their high encoding
complexity, which impacts the cost and power consumption of devices.
Fortunately, the memory and computational cost for encoding LDPC
codes can be greatly reduced by adopting QC-LDPC codes because the
encoder can be implemented by simple shift registers. Furthermore,
IRA LDPC codes are also useful due to their simplicity in designing
and faster encoding than conventional LDPC codes. In addition,
designers can resort to QC-IRA LDPC codes which brings further cost
reduction by combining the features of QC and IRA LDPC codes
\cite{qcirad}, using PEG-based optimization of the graphs and
statistically-driven approaches \cite{baplnc}. These highly
structured LDPC codes could play an important role in the design of
future Wi-Fi standards.

\subsection{Root-Check Codes}

Another family of LDPC codes called Root-Check for block-fading
channels were proposed in \cite{boutros.07}. Root-Check codes are
able to achieve the maximum diversity of a block-fading channel and
have a performance near the limit of outage. Several types of
Root-Check LDPC codes were developed, e.g, \cite{salehi.10,
peg.comms.11,memd}. Among the Root-Check based LDPC codes the ones
designed with the PEG algorithm have shown the best performance
\cite{peg.comms.11}. In particular, we have designed a QC-IRA
Root-Check LDPC code PEG-based with the same block length as that of
the Wi-Fi standard $L = 672$ and with code rate $\frac{1}{2}$. The
key point in this example is to demonstrate by simulations that a
Root-Check based code can significantly outperform the Wi-Fi LDPC
code. For example, consider a simple scenario where the channel is a
block-fading with $F = 2$ fadings, BPSK modulation is used and a
maximum of $20$ decoding iterations are allowed.

%

\begin{figure}[htb]
 \centering
\resizebox{128mm}{!}{
\includegraphics{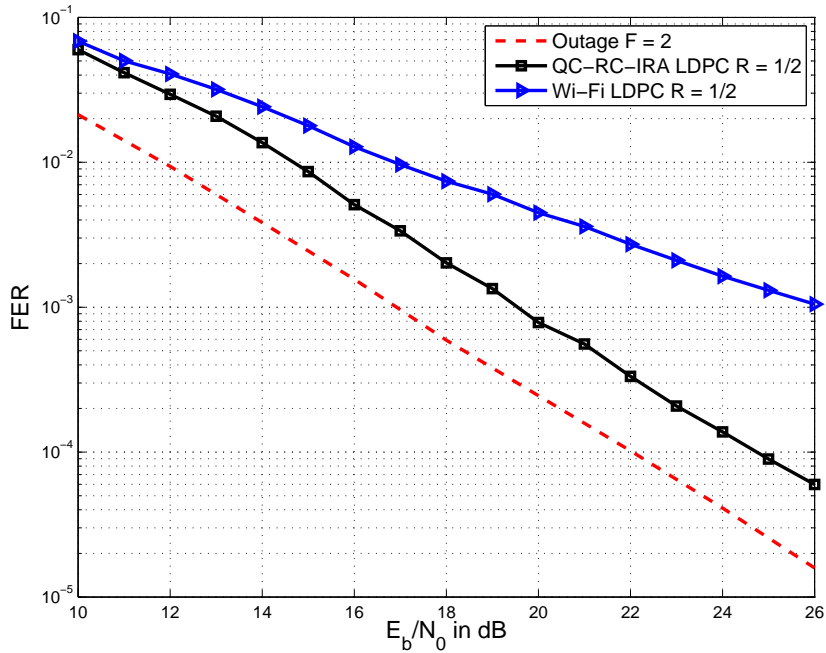}
} \caption{FER performance comparison for QC-IRA Root-Check LDPC
code and Wi-Fi LDPC code over a block-fading channel with $F = 2$
and $L = 672$. The maximum number of iterations is 20. Code rate $R
= \frac{1}{2}$.} \label{fig:wifi_vs_qcrcira}
\end{figure}

In Fig. \ref{fig:wifi_vs_qcrcira} we depict the Frame Error Rate
(FER) performance of a Root-Check LDPC code versus the LDPC code
from Wi-Fi standard both are with the same code rate and with the
same block length. The outage curve is also plotted as reference
which represents the channel capacity.  From Fig.
\ref{fig:wifi_vs_qcrcira}, we can see that the QC-IRA
Root-Check LDPC code outperforms the Wi-Fi LDPC code by $6.5dB$ in
terms of SNR.

\begin{figure}[tb]
 \centering
\resizebox{128mm}{!}{
\includegraphics{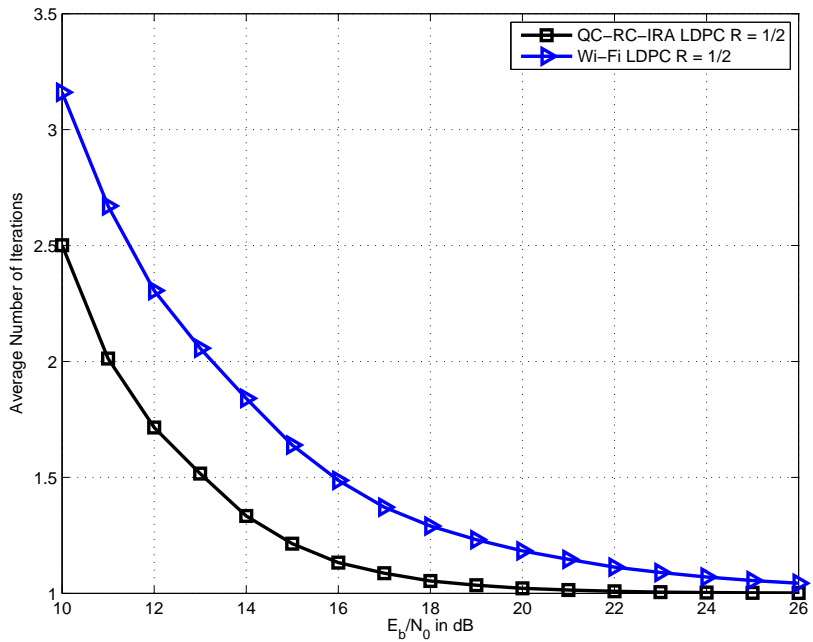}
} \caption{Average number of required iterations for the QC-IRA
Root-Check LDPC code versus the Wi-Fi LDPC code with codeword length $L
= 672$ bits over a block-fading channel with $F = 2$ and code rate $R =
\frac{1}{2}$.} \label{fig:wifi_vs_qcrcira_iter}
\end{figure}

Fig. \ref{fig:wifi_vs_qcrcira_iter} shows the average number of
iterations required by the QC-IRA Root-Check LDPC code versus the Wi-Fi
LDPC code. For the entire SNR region, in average, we observe
that the QC-IRA Root-Check LDPC code requires less decoding
iterations than other LDPC code designs. It must be mentioned that
for medium to high SNR the average required number of iterations is
less than $2$ iterations. The average number of iterations, less
than $2$ at medium to high SNR, corroborates with hardware friendly
LDPC codes. This significantly reduced number of iterations is a key
feature of Root-Check codes, which are limited to rates below $1/2$.
However, other LDPC code designs often require a larger number of
decoding iterations which affect applications sensitive to delay
such as video streaming. For this reason, the development of
advanced decoding algorithms is of utmost importance.

\section{Advanced Decoding Techniques}

In this section, we discuss advanced LDPC decoding techniques which
include reweighting and scheduling approaches, and low-complexity
decoding algorithms. These algorithms offer considerable advantages
over standard belief propagation (BP) algorithms and can address
some of its limitations with regards to performance, delay issues
and computational cost.

\subsection{Reweighting Methods}

The BP algorithm, sometimes also called sum-product algorithm (SPA),
is a powerful algorithm to approximately solve several inference
problems which can be used for error control
coding. Once the BP algorithm was applied as a decoding algorithm
for LDPC codes, various versions of BP graph-based decoding
algorithms have been reported in the area. However, the lack of a
convergence guarantee and the high-latency due to many decoding
iterations are still open issues for researchers when it comes to
effectively decoding LDPC codes in Wi-Fi applications, where such
applications are bandwidth-hungry.

Recently, Wymeersch et al. \cite{urw.bp.detc} introduced the
uniformly reweighted BP (URW-BP) algorithm which exploits BP's
distributed nature and reduces the factor appearance probability
(FAP) to a constant value. In \cite{urw.bp.detc}, the URW-BP has
been shown to outperform the standard BP in terms of LDPC decoding
among other applications. In terms of BP algorithm the URW-BP makes its main modification on the beliefs sent by check nodes to variable nodes. Therefore, the beliefs of the j-$th$ estimated received vector $b(x_{j})$ is:
\begin{equation} \label{eq:urw_update}
b(x_{j}) = \mathcal{L}(x_{j}) + \sum_{i \in \mathcal{N}(j)} \rho \Lambda_{i,j},
\end{equation} where $\mathcal{L}(x_{j})$ is the \textit{a priori} Log Likelihood Ratio (LLR) from the channel, $\rho$ is the constant FAP value, $\Lambda_{i,j}$ are the messages sent from check node $c_i$ to variable node $v_j$ and $i \in \mathcal{N}(j)$ is the neighbouring set of check nodes of $v_j$.

Liu and de Lamare in \cite{vfap.comms} have investigated the idea of
reweighting a suitable part of the factorized graph while also
statistically taking the effect of short cycles into account. By
combining the reweighting strategy with the knowledge of the short
cycles obtained by the cycle counting algorithms they have presented
the variable FAP BP (VFAP-BP) algorithm. The VFAP-BP algorithm
assigns distinct FAP values to each check node on the basis of the
structure of short cycles rather than a complex global graphical
optimization. VFAP-BP outperforms standard algorithms. Furthermore, the VFAP-BP algorithm
can be applied with both regular and irregular LDPC codes while
URW-BP is only advantageous for regular LDPC codes.

\subsection{Scheduling Methods}

The studies in \cite{schedul.wesel,kaids} have suggested that the
use of appropriate scheduling mechanisms for LDPC decoding can
significantly improve the convergence speed in terms of number of
iterations. In general, BP or SPA consist of the exchange of
messages between the nodes of a graph. Each node generates and
propagates messages to its neighbours based on its current incoming
messages.

The LDPC code graph is a bi-partite graph composed by $N_{VN}$
variable nodes $v_{j}$ for $j~\in~\lbrace 1,\cdots, N_{VN}\rbrace$
that represent the codeword bits and $M_{CH}$ check nodes $c_{i}$
for $i~\in~\lbrace 1, \cdots, M_{CH}\rbrace$ that represent the
parity-check equations \cite{schedul.wesel}. In the log-domain
implementations of the BP algorithm, the exchanged messages
correspond to the LLR of the probabilities of the bits. The sign of
the LLR indicates the most likely value of the bit and the absolute
value of the LLR gives the reliability of the message.

BP decoding consists of the iterative update of the messages until a
stopping rule is satisfied. In flooding scheduling, an iteration
consists of the simultaneous update of all the messages from
variable to check nodes followed by the simultaneous update of all
the messages from check to variable nodes. In sequential scheduling,
an iteration consists of the sequential update of all the messages
from variable to check nodes as well as all the messages from check
to variable nodes in a specific pre-defined order. This pre-defined order
is usually designed to allow the parallel processing of the
messages. For instance, the Layered Belief Propagation (LBP) will do
for each check node the update of check to variable nodes followed
by the update of the associated variable nodes. This procedure is
done in an iterative way. The algorithm stops if the decoded bits
satisfy all the parity-check equations or a maximum number of
iterations is reached.

RBP was first introduced in \cite{rbp.2006} and consists of an
informed dynamic scheduling strategy that updates messages according
to an ordering metric called the residual. The message with the
largest residual is updated first. A residual is the norm (defined
over the message space) of the difference between the values of a
message before and after an update. The intuitive justification of
this approach is that as iterative BP converges, the differences
between the messages before and after an update diminish. Then, if a
message has a large residual, it means that it is located in a part
of the graph that has not converged yet. Accordingly, propagating
that message first should speed up convergence.

\begin{figure}[htb]
 \centering
\resizebox{128mm}{!}{
\includegraphics{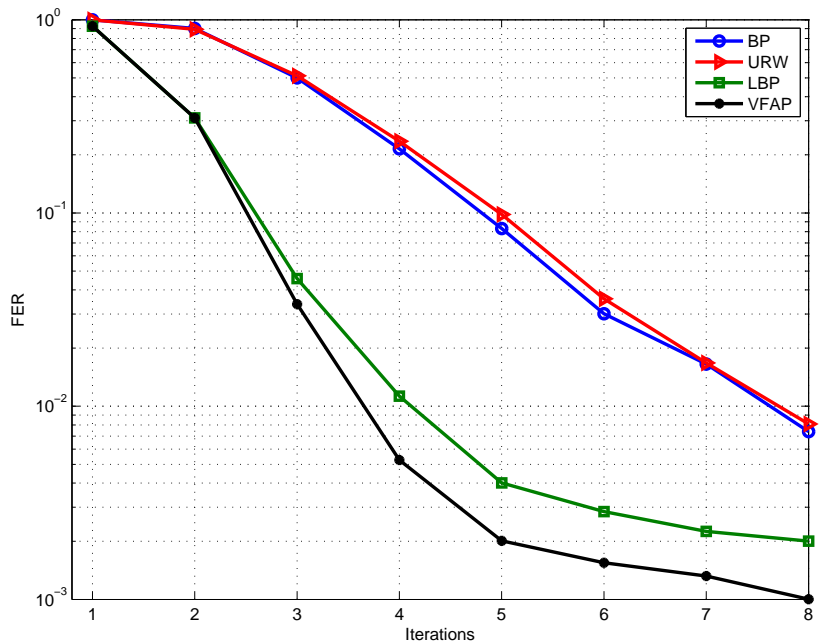}
} \caption{FER performance versus iterations with the Wi-Fi LDPC code rate $R =
\frac{1}{2}$ for BP, URW, LBP and VFAP decoding algorithms. SNR = $3dB$ in the AWGN channel.} \label{fig:fer_iter_r12_3db}
\end{figure}

In order to obtain a better performance, a less greedy scheduling
strategy can be used. The greediness of
RBP comes from the fact that it tends to propagate first the message
to the least reliable node. Vila Casado and et.al. in
\cite{schedul.wesel} proposed to update and propagate simultaneously
all the check-to-variable messages that correspond to the same check
node, instead of only updating and propagating the message with the
largest residual. This algorithm is less likely to propagate the
information from new errors in the next update. This is because
there are many variable nodes that change as opposed to RBP where
only one variable node changes. This strategy is called Node Wise
Scheduling BP (NWBP). Fig. \ref{fig:fer_iter_r12_3db}, depicts an example decoding techniques, which illustrates how they can reduce significantly the number of iterations as compared to BP. With the advent of parallel computing in the latest powerful
processors, the use of scheduling methods can reduce significantly
the overall number of iterations required in the decoding procedure.

With the advent of parallel computing in the latest powerful
processors, the use of scheduling methods can reduce significantly
the overall number of iterations required in the decoding procedure.

\subsection{Reduced Complexity Decoding Methods}
Work on LDPC decoding has mainly focused on floating point
arithmetic or infinite precision (BP algorithm). However, hardware
implementations of decoding algorithms for LDPC codes must address
quantization effects in a fixed-point realization \cite{rcd.ldpc}.
The first approach was to adopt a logarithmic version of the BP
algorithm called Log-BP. Nevertheless, Log-BP sacrifices hardware
implementation due to the fact it requires many hyperbolic tangent
operations.

The Log-BP algorithm can be simplified using the so-called BP-based
approximation (also known as the "min-sum" approximation), which
greatly reduces the implementation complexity, but incurs a
degradation in decoding performance. This has led to the development
of many reduced complexity variants of the BP algorithm that deliver near-optimum decoding performance. There are
three main reduced complexity methods of decoding LDPC codes:
BP-based, Normalized BP-based and Offset BP-based \cite{rcd.ldpc}.

In the min-sum algorithm the key modification is on the horizontal
step or the check to variable node update equation. The check to
variable node update is given by
\begin{equation} \label{eq:min_sum}
\mathcal{L}_{i,j} \approx \left(\prod_{j'\in \mathcal{N}(i)
\setminus j} {\rm sign}(\Lambda_{j',i}) \right) \times
\left(\min_{j' \in \mathcal{N}\setminus j}\vert \Lambda_{j',i}\vert
 \right),
\end{equation} where, $\Lambda_{j',i}$ are the messages sent
from variable node $v_j$ to check node $c_i$ and $j'\in
\mathcal{N}(i) \setminus j$ is the neighbouring set of variable nodes
of $c_i$ except $v_j$. As it can be seen in (\ref{eq:min_sum}) there
is no hyperbolic tangent operation which decreases significantly the
overall computational complexity, although it causes a degradation
in decoding performance.

The min-sum algorithm can be improved by employing a check node
update that uses a normalization constant greater than $\alpha$. The
change made in (\ref{eq:min_sum}) is on the right hand side and
inside the $\min$ operation that is divided by $\alpha$. This method
is called Normalized BP-based. The $\alpha$ parameter should be
adjusted for different SNRs and iterations to achieve its optimum
performance. An effective approach to determine the optimum value of
$\alpha$ is by Density Evolution \cite{rcd.ldpc}.

A computationally more efficient approach that captures the net
effect of the additive correction term applied in each check node
update operation is obtained from the BP-based decoding by
subtracting a positive constant $\beta$. This is called Offset
BP-based. Eq. (\ref{eq:min_sum}) is modified such that $\min_{j' \in
\mathcal{N}\setminus j}\vert \Lambda_{j',i}\vert$ is replaced by
$\max \lbrace \min_{j' \in \mathcal{N}\setminus j}\vert
\Lambda_{j',i}\vert - \beta, 0 \rbrace$. This method differs from
the normalization scheme in that LLR messages smaller in magnitude
than $\beta$ are set to zero, therefore removing their contributions
in the next variable node update step. The min-sum, normalized
BP-based, and Offset BP-based decoding methods do not need any
channel information, e.g., the SNR, and work with just the received
values as inputs. Both offset and
normalized BP-based decoding algorithms can achieve performance very
close to that of BP decoding while offering significant advantages
for hardware implementation.


\section{Challenges}

The LDPC codes used in the Wi-Fi standard are not optimally designed. Therefore, it is important to employ design
algorithms capable of producing LDPC codes of high performance
and efficiency for short to moderate lengths such as PEG-based
algorithms. The example shown in Fig. \ref{fig:wifi_vs_qcrcira} poses the following
question: Why are the Wi-Fi LDPC codes not optimally or near optimally designed? We consider this as an open problem for future Wi-Fi
standards.

The PEG-based QC-IRA Root-Check LDPC code designed was compared to
the Wi-Fi LDPC code to analyse the overall performance in terms of
FER. As seen in Fig. \ref{fig:wifi_vs_qcrcira} the PEG-based
designed code has outperformed the Wi-Fi LDPC code by about $6.5dB$.
Therefore, future Wi-Fi standards would have benefits if they
considered the use of Root-Check based LDPC codes for rate $R =
\frac{1}{2}$ scenarios.

The use of scheduling methods to decode LDPC codes can reduce
significantly the number of iterations required in the
decoding procedure. Accordingly, the resulting number of iterations
can increase the advantages LDPC codes over convolutional codes,
which are widely used in portable devices due to their simplicity.
An open problem in LDPC decoding is: Which is the most
cost-effective method for the Wi-Fi standard? This is an
investigation that will keep designers of Wi-Fi systems busy in the
next few years.

In terms of adoption of LDPC codes as FEC in Wi-Fi standards, key
issues are the encoding and decoding operations. The cost of
encoding can be decreased significantly by using RA based LDPC codes
which are able to provide simple and low complexity encoding
methods. The decoding cost is another key issue that must be addressed.
One way to addressing such problem is by introducing reduced complexity
decoding (RCD) algorithms, e.g., min-sum, normalized BP-based and
Offset BP-based. The choice of decoding algorithm is another
important design challenge, which must take into account aspects
such low complexity operations, stability and performance.

\section{Conclusion}
This article has discussed coding techniques for Wi-Fi systems. In
particular, we have discussed approaches to reducing the
complexity of encoding and decoding and improving the design and the
decoding of LDPC codes. The main challenge is in decoding which is how to obtain an attractive trade-off between performance and computational complexity. To conclude, we advocate that future Wi-Fi standards should use LDPC codes as the main FEC to fulfil the needs of bandwidth-hungry and high-performance applications.


\section*{Acknowledgment}
This work was partially supported by PNPD/CAPES (Brazil).

\bibliographystyle{IEEEtran}
%

%
\bibliography{IEEEabrv,coding_wifi}

\end{document}